\documentclass[prapplied,twocolumn,showpacs,preprintnumbers,amsmath,amssymb,superscriptaddress]{revtex4-1}

\usepackage{graphicx}
\usepackage{dcolumn}
\usepackage{bm}
\usepackage{units}

\begin{document}

\title{Spin structure of textured and isotropic Nd-Fe-B-based nanocomposites: evidence for correlated crystallographic and spin texture}

\author{A.~Michels}
\email[]{andreas.michels@uni.lu}
\author{R.~Weber}
\author{I.~Titov}
\author{D.~Mettus}
\affiliation{Physics and Materials Science Research Unit, University of Luxembourg, 162a~avenue de la Faïencerie, L-1511~Luxembourg, Luxembourg}
\author{$\mathrm{\acute{E}}$.A.~P{\'e}rigo}
\affiliation{Physics and Materials Science Research Unit, University of Luxembourg, 162a~avenue de la Faïencerie, L-1511~Luxembourg, Luxembourg}
\affiliation{ABB Corporate Research Center, 940~Main Campus Drive, 27606~Raleigh, North Carolina}
\author{I.~Peral}
\affiliation{Physics and Materials Science Research Unit, University of Luxembourg, 162a~avenue de la Faïencerie, L-1511~Luxembourg, Luxembourg}
\affiliation{Materials Research and Technology Department, Luxembourg Institute of Science and Technology, 41~rue du Brill, L-4422 Belvaux, Luxembourg}
\author{O. Vallcorba}
\affiliation{Alba Synchrotron, BP~1413, km~3.3, Cerdanyola del Vall$\grave{e}$s, Spain}
\author{J.~Kohlbrecher}
\affiliation{Paul Scherrer Institute, CH-5232~Villigen PSI, Switzerland}
\author{K.~Suzuki}
\affiliation{Department of Materials Science and Engineering, Monash University, Clayton, Victoria~3800, Australia}
\author{M.~Ito}
\author{A.~Kato}
\author{M.~Yano}
\affiliation{Advanced Material Engineering Division, Toyota Motor Corporation, Susono 410-1193, Japan}

\date{\today}

\begin{abstract}
We report the results of a comparative study of the magnetic microstructure of textured and isotropic $\mathrm{Nd}_2\mathrm{Fe}_{14}\mathrm{B}/\alpha$-$\mathrm{Fe}$ nanocomposites using magnetometry, transmission electron microscopy, synchrotron x-ray diffraction, and, in particular, magnetic small-angle neutron scattering (SANS). Analysis of the magnetic neutron data of the textured specimen and computation of the correlation function of the spin misalignment SANS cross section suggests the existence of inhomogeneously magnetized regions on an intraparticle nanometer length scale, about $40-50 \, \mathrm{nm}$ in the remanent state. Possible origins for this spin disorder are discussed: it may originate in thin grain-boundary layers (where the materials parameters are different than in the $\mathrm{Nd}_2\mathrm{Fe}_{14}\mathrm{B}$ grains), or it may reflect the presence of crystal defects (introduced via hot pressing), or the dispersion in the orientation distribution of the magnetocrystalline anisotropy axes of the $\mathrm{Nd}_2\mathrm{Fe}_{14}\mathrm{B}$ grains. X-ray powder diffraction data reveal a crystallographic texture in the direction perpendicular to the pressing direction -- a finding which might be related to the presence of a texture in the magnetization distribution, as inferred from the magnetic SANS data.
\end{abstract}


\maketitle

\section{Introduction}

Nd-Fe-B-based nanocomposite permanent magnets, which consist of exchange-coupled nanocrystalline hard ($\mathrm{Nd}_2\mathrm{Fe}_{14}\mathrm{B}$) and soft ($\alpha$-$\mathrm{Fe}$ or $\mathrm{Fe}_3\mathrm{B}$) magnetic phases, are of potential interest for electronic devices due to their preeminent magnetic properties such as high remanence and magnetic energy product \cite{gutfleisch2011,liu2009}. The major challenge remains the understanding of how the details of the microstructure (e.g., average particle size and shape, volume fraction of soft phase, texture, interfacial chemistry) correlate with their magnetic properties. In order to tackle this issue a multiscale characterization approach is adopted, which comprises a suite of both experimental and theoretical state-of-the-art methods such as high-resolution electron microscopy, electron backscattering diffraction, three-dimensional atom-probe analysis, Lorentz and Kerr microscopy, or atomistic and continuum micromagnetic simulations.

Recent investigations by Liu \textit{et al.} \cite{liu2013} and Sepehri-Amin \textit{et al.} \cite{SepehriAmin2013} demonstrate that the properties of the interface regions between the $\mathrm{Nd}_2\mathrm{Fe}_{14}\mathrm{B}$ grains decisively determine the coercivity of the sample. The grain-boundary layers (and triple-junctions between the grains) have a thickness between about $1-15 \, \mathrm{nm}$ and can be both in a crystalline or amorphous state. Moreover, as far as their magnetism is concerned, the intergranular regions are characterized by different magnetic interactions (exchange, magnetocrystalline anisotropy, saturation magnetization) as compared to the $\mathrm{Nd}_2\mathrm{Fe}_{14}\mathrm{B}$ crystallites and, hence, they represent potential sources for the nucleation of inhomogeneous spin textures during magnetization reversal. Indeed, the micromagnetic simulation results reported in \cite{SepehriAmin2013} suggest that the existence of a thin ($< 5 \, \mathrm{nm}$) ferromagnetic grain-boundary phase with reduced magnetocrystalline anisotropy, exchange-stiffness constant, and saturation magnetization causes the magnetization reversal to occur from the soft intergranular phase into the hard $\mathrm{Nd}_2\mathrm{Fe}_{14}\mathrm{B}$ phase at a field of $-2.5 \, \mathrm{T}$. When the intergranular phase is nonmagnetic, then the nucleation of reversed domains starts from the triple junctions of the $\mathrm{Nd}_2\mathrm{Fe}_{14}\mathrm{B}$ grains at a higher field of $-3.2 \, \mathrm{T}$.

In this context it also worth noting that first-principles density functional theory calculations on an exchange-spring multilayer system \cite{toga2016} predict a dependency of the exchange coupling on the crystallographic orientation at the interface between $\mathrm{Nd}_2\mathrm{Fe}_{14}\mathrm{B}$ and $\alpha$-$\mathrm{Fe}$; specifically, ferromagnetic coupling is predicted for the $\mathrm{Nd}_2\mathrm{Fe}_{14}\mathrm{B}(001)/$$\alpha$-$\mathrm{Fe}(001)$ interface model, whereas antiferromagnetic interactions are obtained for $\mathrm{Nd}_2\mathrm{Fe}_{14}\mathrm{B}(100)/$$\alpha$-$\mathrm{Fe}(110)$. If this prediction were true, then it may negatively influence the magnetic properties of this class of materials (e.g., the maximum energy product). Indeed, a recent experimental study using ferromagnetic resonance and Kerr microscopy \cite{ogawa2015} reports on the predicted negative exchange coupling.

The above discussed examples ultimately demonstrate that the magnetic microstructure of nanocrystalline Nd-Fe-B-based magnets is characterized by inhomogeneous magnetization structures and that interfacial regions are a major cause for the nanoscale spin disorder. In addition to the grain boundaries, there exist, however, other sources of spin disorder in such materials: ultrafine-grained textured nanocomposites are produced from melt-spun ribbons via hot compaction \cite{gutfleisch2000,kirchner2004,liu2013,SepehriAmin2013}; this process may introduce crystal defects which locally act as nucleation centers for nonuniform magnetization textures. Furthermore, one has to invoke a magnetization inhomogeneity which is due to the nonideal alignment (dispersion) of the crystallographic $c$-axes (of the $\mathrm{Nd}_2\mathrm{Fe}_{14}\mathrm{B}$ grains) along the pressing direction during hot deformation; the spins have to undergo rotations in order to accommodate to the changes in the easy-axis magnetization direction from grain to grain. Last but not least, there is the magnetic shape anisotropy of the usually platelet-shaped $\mathrm{Nd}_2\mathrm{Fe}_{14}\mathrm{B}$ particles, which may result in a small spin canting towards the plane perpendicular to the $c$-axis. It is certainly true that the magnetic anisotropy field of the $\mathrm{Nd}_2\mathrm{Fe}_{14}\mathrm{B}$ phase (about $8 \, \mathrm{T}$ at $300 \, \mathrm{K}$ \cite{woodcock2012}) is much larger than any shape-anisotropy field (assuming, e.g., $0.5 \, \mathrm{T}$ for strongly anisotropic grains), but nevertheless weak spin canting ($\tan^{-1}(0.5/8) \cong 3.6^{\circ}$) might be produced by the competition between shape and magnetocrystalline anisotropy.

In order to scrutinize the above-sketched issue, we have carried out a comparative study of the magnetic microstructure of textured (hot-deformed) and isotropic nanocrystalline $\mathrm{Nd}_2\mathrm{Fe}_{14}\mathrm{B}/\alpha$-$\mathrm{Fe}$ by means of magnetic small-angle neutron scattering (SANS). Specifically, the central aim of our investigation is to detect and quantify the presumed nanoscale spin disorder, which is commonly only indirectly inferred by combining results from electron microscopy, magnetization, and micromagnetic simulations. 

The SANS technique (see Ref.~\cite{michels2014review} for a review) provides information on variations of both the magnitude and orientation of the magnetization on a nanometer length scale ($\sim 1-300$~nm). SANS is extremely sensitive to long-wavelength magnetization fluctuations and it has only recently been employed for characterizing Nd-Fe-B-based permanent magnets: for example, the field dependence of characteristic magnetic length scales during the magnetization-reversal process in isotropic Nd-Fe-B-based nanocomposites \cite{bickapl2013} and in isotropic sintered Nd-Fe-B \cite{perigo2015} was studied, the exchange-stiffness constant has been determined \cite{bickapl2013no2}, the observation of the so-called spike anisotropy in the magnetic SANS cross section has been explained with the formation of flux-closure patterns \cite{perigo2014}, magnetic multiple scattering has been detected \cite{ono2016}, textured Nd-Fe-B has been investigated \cite{perigo2016no1}, and the effect of grain-boundary diffusion on the magnetization-reversal process of isotropic \cite{perigo2016no2} and hot-deformed textured \cite{yano2012,yano2014,saito2015} nanocrystalline Nd-Fe-B magnets has been studied.

\section{Experimental}

Two $\mathrm{Nd}_2\mathrm{Fe}_{14}\mathrm{B}/\alpha$-$\mathrm{Fe}$ nanocomposites containing, respectively, $5 \, \mathrm{wt\%}$ of Fe were investigated in this study. Both samples were prepared by means of the melt-spinning technique. One sample was subsequently hot-deformed in order to obtain a textured magnet. For this purpose, the melt-spun ribbons were crushed into powders of a few hundred micrometers in size and then sintered at $973 \, \mathrm{K}$ under a pressure of $400 \, \mathrm{MPa}$. The sintered bulk was hot-deformed with a height reduction of about $75 \, \%$ to develop the [001] texture of the $\mathrm{Nd}_2\mathrm{Fe}_{14}\mathrm{B}$ phase along the pressing direction \cite{kirchner2004,SepehriAmin2013}. This results in the formation of platelet-shaped $\mathrm{Nd}_2\mathrm{Fe}_{14}\mathrm{B}$ grains with an average thickness of $\sim 110 \, \mathrm{nm}$ and an average diameter of $\sim 140 \, \mathrm{nm}$. The $\mathrm{Nd}_2\mathrm{Fe}_{14}\mathrm{B}$ platelets are stacked along the nominal $c$-axis, which we define as the [001] direction, with some degrees of misorientation. The isotropic sample had an average grain size of about $20 \, \mathrm{nm}$. We have also investigated composites with $0 \, \mathrm{wt\%}$ and $10 \, \mathrm{wt\%}$ of Fe, which, as far as the neutron results are concerned, show qualitatively the same behavior as the $5 \, \mathrm{wt\%}$ sample. For further details, see Refs.~\cite{yano2012,yano2014,saito2015}.

\begin{figure}[htb!]
\includegraphics[width=1.0\columnwidth]{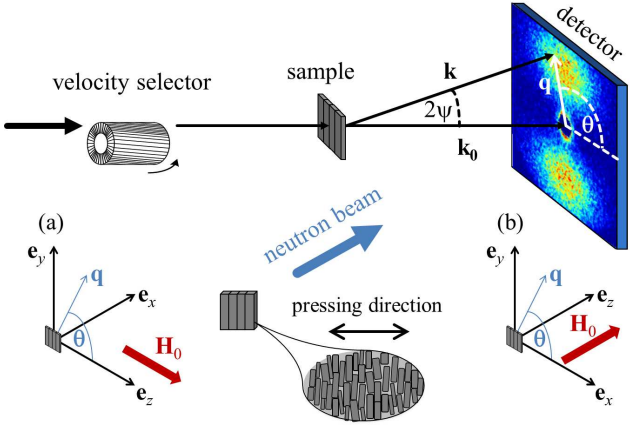}
\caption{\label{fig1} Sketch of the perpendicular (a) and parallel (b) scattering geometry, which, respectively, have the applied magnetic field $\mathbf{H}_0$ perpendicular and parallel to the wave vector $\mathbf{k}_0$ of the incident neutron beam; $q = \left|\mathbf{q}\right| = 4\pi \lambda^{-1} \sin\psi$, where $2\psi$ denotes the scattering angle and $\lambda$ is the mean neutron wavelength. Note that $\mathbf{H}_0 \parallel \mathbf{e}_z$ in \textit{both} geometries and that $\mathbf{q} \cong (0, q_y, q_z) = q (0, \sin\theta, \cos\theta)$ for $\mathbf{k}_0 \perp \mathbf{H}_0$ (a) and $\mathbf{q} \cong (q_x, q_y, 0) = q (\cos\theta, \sin\theta, 0)$ for $\mathbf{k}_0 \parallel \mathbf{H}_0$ (b). The pressing direction is horizontal, which is along $\mathbf{e}_z$ in (a) and along $\mathbf{e}_x$ in (b).}
\end{figure}

The neutron experiment has been carried out at $300 \, \mathrm{K}$ at the instrument SANS-I at the Paul Scherrer Institute, Switzerland, using unpolarized neutrons with a mean wavelength of $\lambda = 4.5 \, \mathrm{\AA}$ and $\Delta \lambda / \lambda = 10 \, \mathrm{\%}$ (FWHM) \cite{kohlbrecher2000,niketic2015}. The external magnetic field $\mathbf{H}_0$ (provided by a cryomagnet; $\mu_0 H_{\mathrm{max}} = 9.5 \, \mathrm{T}$) was applied perpendicular and parallel to the wave vector $\mathbf{k}_0$ of the incoming neutron beam (compare Fig.~\ref{fig1}); this corresponds to the situation that $\mathbf{H}_0$ is parallel ($\mathbf{k}_0 \perp \mathbf{H}_0$) and perpendicular ($\mathbf{k}_0 \parallel \mathbf{H}_0$) to the nominal $c$-axis (pressing direction) of the textured sample. Neutron data were corrected for background scattering (empty sample holder), transmission, and detector efficiency using the GRASP software package. The measured transmission was larger than $90 \, \mathrm{\%}$ for both samples at all fields investigated. Further sample characterization was done by means of vibrating sample magnetometry, transmission electron microscopy, and synchrotron x-ray diffraction (at beamline BL04-MSPD at the Alba synchrotron, Barcelona, Spain \cite{Fauth2015}).

\section{Unpolarized SANS cross sections and correlation function}

The elastic unpolarized SANS cross section $d \Sigma / d \Omega$ at momentum-transfer vector $\mathbf{q}$ takes on different forms depending on the relative orientation between the wave vector $\mathbf{k}_0$ of the incident neutron beam and the externally applied magnetic field $\mathbf{H}_0$ \cite{michels2014review}; for the perpendicular geometry  ($\mathbf{k}_0 \perp \mathbf{H}_0$), we obtain
\begin{eqnarray}
\label{sigmatotperp}
\frac{d \Sigma}{d \Omega}(\mathbf{q}) = \frac{8 \pi^3}{V} \left( |\widetilde{N}|^2 + b_H^2 |\widetilde{M}_x|^2 + b_H^2 |\widetilde{M}_y|^2 \cos^2\theta \right. \nonumber \\ \left. + b_H^2 |\widetilde{M}_z|^2 \sin^2\theta - b_H^2 (\widetilde{M}_y \widetilde{M}_z^{\ast} + \widetilde{M}_y^{\ast} \widetilde{M}_z) \sin\theta \cos\theta \right) , \nonumber \\
\end{eqnarray}
whereas for the parallel case ($\mathbf{k}_0 \parallel \mathbf{H}_0$)
\begin{eqnarray}
\label{sigmatotpara}
\frac{d \Sigma}{d \Omega}(\mathbf{q}) = \frac{8 \pi^3}{V} \left( |\widetilde{N}|^2 + b_H^2 |\widetilde{M}_x|^2 \sin^2\theta + b_H^2 |\widetilde{M}_y|^2 \cos^2\theta \right. \nonumber \\ \left. + b_H^2 |\widetilde{M}_z|^2 - b_H^2 (\widetilde{M}_x \widetilde{M}_y^{\ast} + \widetilde{M}_x^{\ast} \widetilde{M}_y) \sin\theta \cos\theta \right) ; \nonumber \\
\end{eqnarray}
$V$ denotes the scattering volume, $b_H = 2.91 \times 10^{8} \, \mathrm{A^{-1} m^{-1}}$, $\widetilde{N}(\mathbf{q})$ is the nuclear scattering amplitude, and $\widetilde{\mathbf{M}}(\mathbf{q}) = \{ \widetilde{M}_x(\mathbf{q}), \widetilde{M}_y(\mathbf{q}), \widetilde{M}_z(\mathbf{q}) \}$ represents the Fourier transform of the magnetization $\mathbf{M}(\mathbf{r}) = \{ M_x(\mathbf{r}), M_y(\mathbf{r}), M_z(\mathbf{r}) \}$; $c^{*}$ is a quantity complex-conjugated to $c$. We would like to emphasize that the magnetization vector field of a bulk ferromagnet is a function of the position $\mathbf{r} = \{ x, y, z \}$ inside the material, i.e., $\mathbf{M} = \mathbf{M}(x, y, z)$, and that, consequently, $\widetilde{\mathbf{M}} = \widetilde{\mathbf{M}}(q_x, q_y, q_z)$. However, the Fourier components which appear in the above SANS cross sections represent projections into the $q_y$-$q_z$-plane for $\mathbf{k}_0 \perp \mathbf{H}_0$ ($q_x \cong 0$) and into the $q_x$-$q_y$-plane for $\mathbf{k}_0 \parallel \mathbf{H}_0$ ($q_z \cong 0$) (compare Fig.~\ref{fig1}). In polar coordinates, the $\widetilde{M}_{x,y,z}$ then depend (in addition to the applied field and the magnetic interactions) on both the magnitude $q$ and the orientation $\theta$ of the scattering vector $\mathbf{q}$ \cite{erokhin2015}. 

\begin{figure*}[tb!]
\includegraphics[width=1.65\columnwidth]{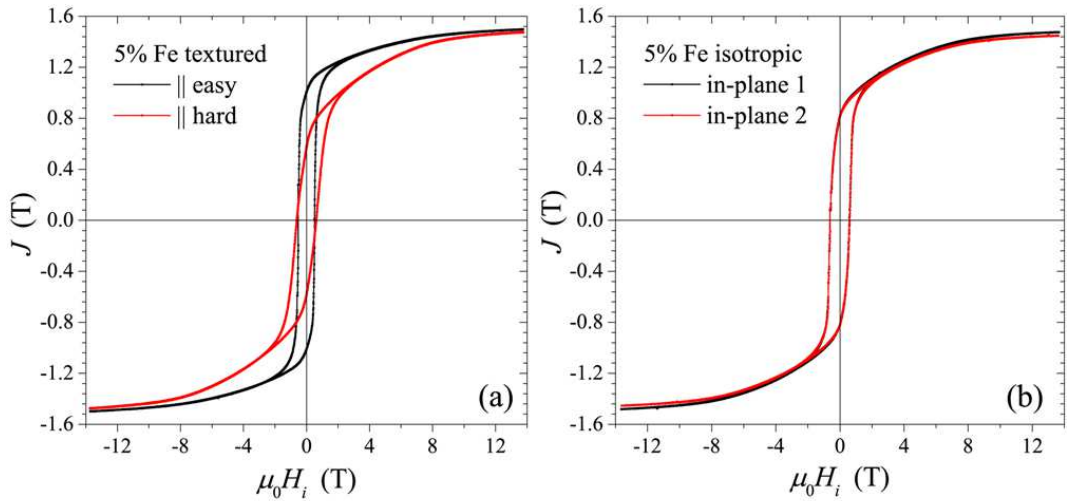}
\caption{\label{fig2} Room-temperature magnetization curves of textured (a) and isotropic (b) $\mathrm{Nd}_2\mathrm{Fe}_{14}\mathrm{B}/\alpha$-$\mathrm{Fe}$ ($5 \, \mathrm{wt\%}$ Fe). Measurements have been carried out for the magnetic field applied parallel and perpendicular to the texture axis (pressing direction) in (a), and for two different in-plane directions in (b) (``in-plane~2'' direction is rotated by $90^{\circ}$ with respect to ``in-plane~1'' direction). Magnetization data (on the rectangular-shaped samples) have been corrected for demagnetizing effects using the magnetometric demagnetizing factor \cite{aharoni98}.}
\end{figure*}

In our neutron data analysis below, we subtract the respective SANS signal at the largest available field of $9.5 \,  \mathrm{T}$ (approach-to-saturation regime, compare Fig.~\ref{fig2}) from the measured data at lower fields. This subtraction procedure eliminates the nuclear SANS contribution ($\propto |\widetilde{N}|^2$), which is field independent, and it yields the so-called spin-misalignment SANS cross section $d \Sigma_M / d \Omega$, which we display here for simplicity only for the parallel scattering geometry:
\begin{eqnarray}
\label{sigmasmpara}
\frac{d \Sigma_M}{d \Omega} = \frac{8 \pi^3}{V} \, b_H^2 \left( \Delta |\widetilde{M}_x|^2 \sin^2\theta + \Delta |\widetilde{M}_y|^2 \cos^2\theta \right. \nonumber \\ \left. + \Delta |\widetilde{M}_z|^2 + \Delta CT \sin\theta \cos\theta \right) ,
\end{eqnarray}
where $\Delta |\widetilde{M}_x|^2 := |\widetilde{M}_x|^2(H) - |\widetilde{M}_x|^2(9.5 \,  \mathrm{T})$ (and so on for the other Fourier coefficients) represents the difference between the value of $|\widetilde{M}_x|^2$ at the actual field $H$ and the measurement at $9.5 \,  \mathrm{T}$ [$CT := - (\widetilde{M}_x \widetilde{M}_y^{\ast} + \widetilde{M}_x^{\ast} \widetilde{M}_y$)]. If it would be possible to fully saturate the sample (i.e., $\mathbf{M}(\mathbf{r}) = \{ 0, 0, M_z = M_s(\mathbf{r}) \}$) and if one restricts the considerations (subtraction procedure) to the approach-to-saturation regime, where the field dependence of the longitudinal Fourier component can be neglected (i.e., $|\widetilde{M}_z|^2(H) - |\widetilde{M}_s|^2 \rightarrow 0$), then $d \Sigma_M / d \Omega$ (for $\mathbf{k}_0 \parallel \mathbf{H}_0$) reduces to
\begin{eqnarray}
\label{sigmasmparainfinity}
\frac{d \Sigma_M}{d \Omega} = \frac{8 \pi^3}{V} \, b_H^2 \left( |\widetilde{M}_x|^2 \sin^2\theta + |\widetilde{M}_y|^2 \cos^2\theta \right. \nonumber \\ \left. + CT \sin\theta \cos\theta \right) ,
\end{eqnarray}
and likewise for the $\mathbf{k}_0 \perp \mathbf{H}_0$ geometry.

Furthermore, it is decisive for the later discussion to note that for a ferromagnet with a statistically-isotropic microstructure the parallel total (nuclear and magnetic) $d \Sigma / d \Omega$ and $d \Sigma_M / d \Omega$ [Eqs.~(\ref{sigmatotpara}) and (\ref{sigmasmpara})] are generally \textit{isotropic}, i.e., $\theta$-independent (see, e.g., Fig.~21 in Ref.~\cite{michels2014review}, Fig.~4 in Ref.~\cite{michels2014jmmm}, or Figs.~\ref{fig5}(b), \ref{fig7}(c), and \ref{fig7}(d) below). In other words, although the individual contributions to the parallel SANS cross section are highly anisotropic (e.g., $|\widetilde{M}_x|^2 \sin^2\theta$), their corresponding sums in Eqs.~(\ref{sigmatotpara}) and (\ref{sigmasmpara}) are isotropic for a statistically-isotropic ferromagnet; this is not true for the perpendicular geometry [Eq.~(\ref{sigmatotperp})], which generally exhibits a pronounced angular anisotropy.

The (normalized) correlation function $c(r)$ of the spin misalignment can be computed from azimuthally-averaged data via \cite{mettus2015}
\begin{equation}
\label{corrfunc}
c(r) = \frac{\int_0^{\infty} \frac{d \Sigma_M}{d \Omega}(q) J_0(q r) \, q \, dq}{\int_0^{\infty} \frac{d \Sigma_M}{d \Omega}(q) \, q \, dq} ,
\end{equation}
where $J_0(qr)$ denotes the zeroth-order Bessel function. Analysis of $c(r)$ provides information on the characteristic magnetic length scales \cite{bickapl2013,perigo2015,perigo2016no2}.

\begin{figure*}[tb!]
\centering
\includegraphics[width=1.90\columnwidth]{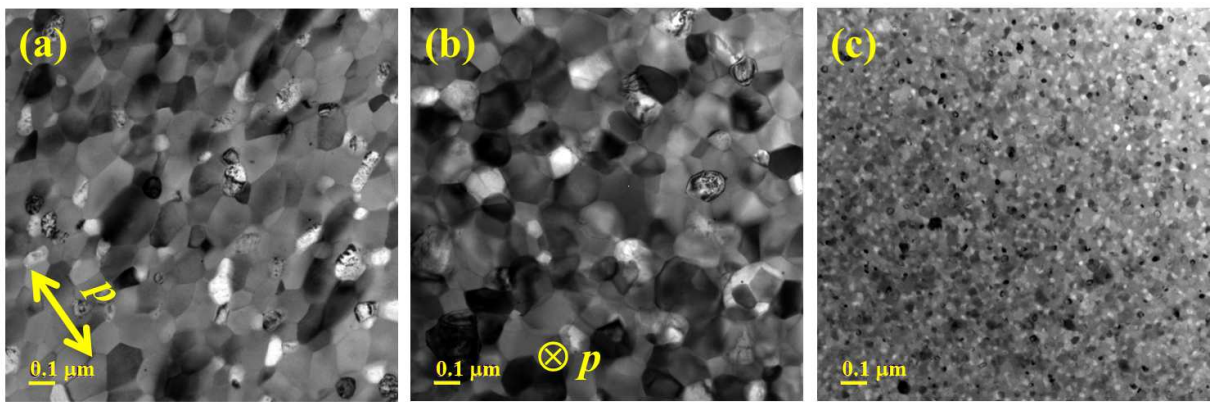}
\caption{\label{fig3} Bright-field transmission electron microscopy images of the textured [(a) and (b)] and isotropic (c) $\mathrm{Nd}_2\mathrm{Fe}_{14}\mathrm{B}/\alpha$-$\mathrm{Fe}$ nanocomposites ($5 \, \mathrm{wt\%}$ Fe). The average sizes of the anisotropic grains of the textured sample have been estimated [from (a) and (b)] as, respectively, $\sim 110 \, \mathrm{nm}$ (parallel to the pressing direction ``$p$'') and $\sim 140 \, \mathrm{nm}$ (perpendicular to the pressing direction ``$p$''), while the average grain diameter of the isotropic sample has been found to be $\sim 20 \, \mathrm{nm}$.}
\end{figure*}

\begin{figure*}[htb!]
\centering
\includegraphics[width=2.0\columnwidth]{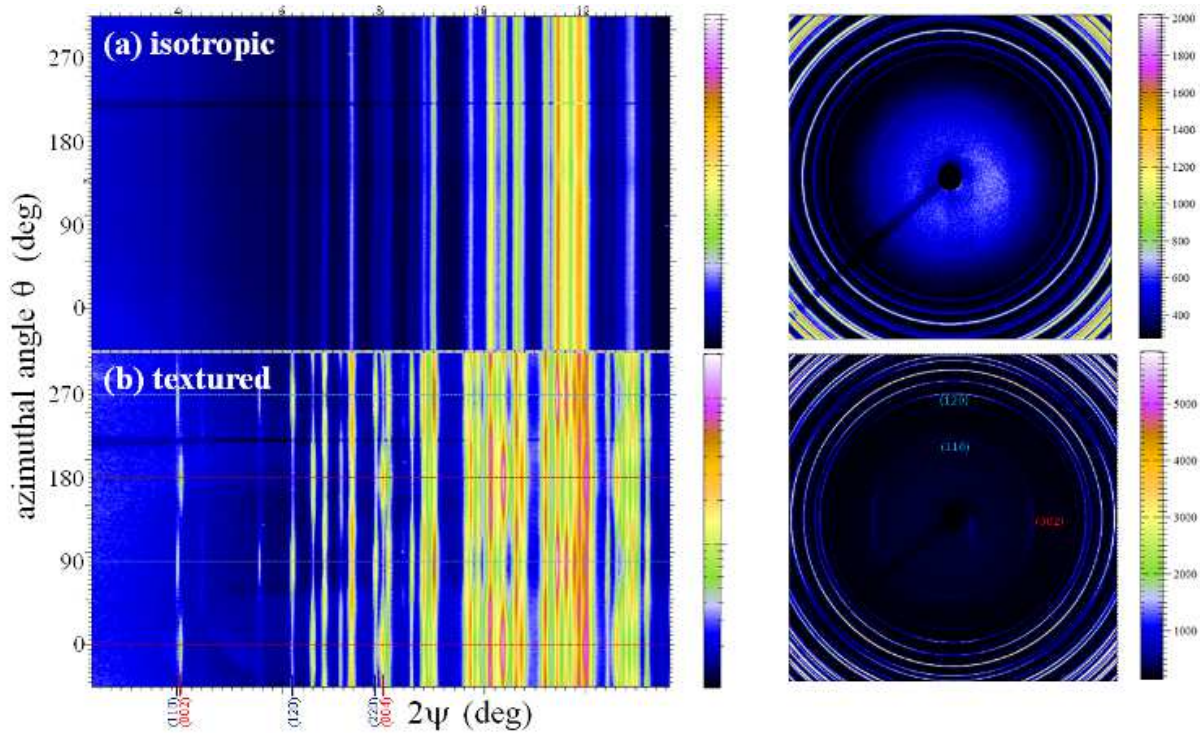}
\caption{\label{fig4} Synchrotron x-ray diffraction data of the isotropic (a) and textured (b) $\mathrm{Nd}_2\mathrm{Fe}_{14}\mathrm{B}/\alpha$-$\mathrm{Fe}$ nanocomposites ($5 \, \mathrm{wt\%}$ Fe). The pressing direction of the hot-deformed sample is horizontal (same scattering geometry as in the neutron experiment, compare Fig.~\ref{fig1}). (left images) Integrated intensity as a function of azimuthal angle $\theta$ and scattering angle $2 \psi$; (right images) corresponding Debye-Scherrer diffraction rings. Radial integration of synchrotron data has been performed with the Fit2D software \cite{fit2d}.}
\end{figure*}

\section{Results and discussion}

Figure~\ref{fig2} displays the room-temperature magnetization curves of textured [Fig.~\ref{fig2}(a)] and isotropic [Fig.~\ref{fig2}(b)] $\mathrm{Nd}_2\mathrm{Fe}_{14}\mathrm{B}/\alpha$-$\mathrm{Fe}$. The coercive fields are $\mu_0 H_c = 0.57 \, \mathrm{T}$ (textured) and $\mu_0 H_c = 0.61 \, \mathrm{T}$ (isotropic). The saturation polarization was estimated by extrapolating the data to infinite field: we find $J_s = 1.57 \, \mathrm{T}$ (textured) and $J_s = 1.53 \, \mathrm{T}$ (isotropic) with ensuing remanence-to-saturation ratios of about $0.64$ (textured easy), $0.36$ (textured hard), and $0.53$ (isotropic). Consistent with the magnetization data, the transmission electron microscopy images of the textured sample [Fig.~\ref{fig3}(a) and \ref{fig3}(b)] reveal a weakly anisotropic microstructure, while the isotropic sample [Fig.~\ref{fig3}(c)] exhibits equiaxed grains.

X-ray diffraction measurements carried out (in transmission mode) at the Alba synchrotron (Fig.~\ref{fig4}) unambiguously prove the presence of a weak texture along the (horizontal) pressing direction; specifically, diffraction peaks of the type $(0 0 l)$ do present two maxima around $\theta = 0^{\circ}$ and $\theta = 180^{\circ}$ in the Debye Scherrer rings. Additionally, we find evidence for the presence of texture along other crystallographic directions; peaks of the type $(h k 0)$ exhibit two maxima around $90^{\circ}$ and $270^{\circ}$. This latter observation will be of relevance when discussing the results of the magnetic neutron data analysis (see below).

Figure~\ref{fig5} depicts (for $\mathbf{k}_0 \parallel \mathbf{H}_0$) the two-dimensional unpolarized total scattering cross sections $d \Sigma / d \Omega$ of the textured and isotropic Nd-Fe-B-based nanocomposites at selected applied magnetic fields ($9.5 \,  \mathrm{T}$, remanence, coercive field). The isotropic sample [Fig.~\ref{fig5}(b)] exhibits an isotropic scattering pattern at all fields investigated, whereas the textured sample [Fig.~\ref{fig5}(a)] shows anisotropic scattering with an elongation along the horizontal direction. The corresponding (over $2\pi$) azimuthally-averaged data sets are displayed in Fig.~\ref{fig6}; between the coercive field and the largest available field of $9.5 \,  \mathrm{T}$, the cross section of the isotropic sample changes (roughly) by about an order of magnitude at the smallest momentum-transfers $q$ (and about half an order of magnitude for the textured sample).

\begin{figure*}[tb!]
\includegraphics[width=1.90\columnwidth]{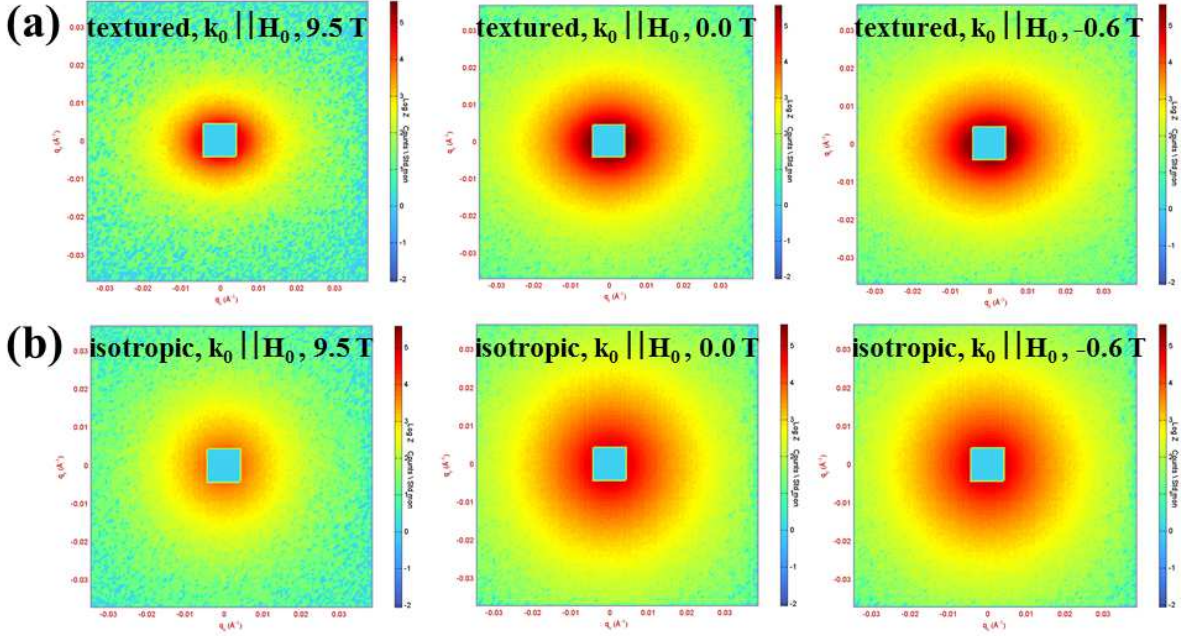}
\caption{\label{fig5} Color-coded two-dimensional intensity maps of the total unpolarized $d \Sigma / d \Omega$ in the plane perpendicular to the incoming neutron beam at selected applied magnetic fields (see insets) (logarithmic color scale) ($\mathbf{k}_0 \parallel \mathbf{H}_0$). $d \Sigma / d \Omega$ of the textured (a) and isotropic (b) $\mathrm{Nd}_2\mathrm{Fe}_{14}\mathrm{B}/\alpha$-$\mathrm{Fe}$ nanocomposite. $\mathbf{H}_0$ is normal to the detector plane.}
\end{figure*}
\begin{figure*}[tb!]
\includegraphics[width=2.0\columnwidth]{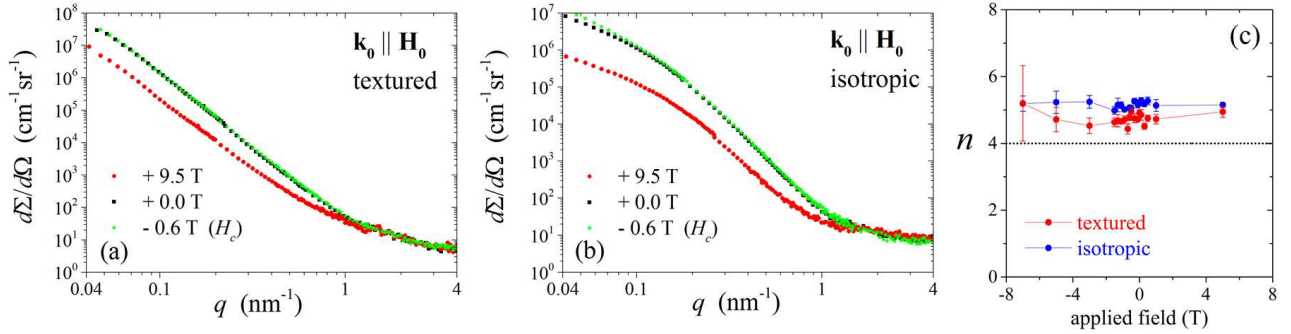}
\caption{\label{fig6} Azimuthally-averaged total unpolarized SANS cross sections $d \Sigma / d \Omega$ at selected applied magnetic fields (see insets) (log-log scale) ($\mathbf{k}_0 \parallel \mathbf{H}_0$). $d \Sigma / d \Omega$ of the textured (a) and isotropic (b) $\mathrm{Nd}_2\mathrm{Fe}_{14}\mathrm{B}/\alpha$-$\mathrm{Fe}$ nanocomposite. (c) Applied-field dependence of the power-law exponent $n$ in $d \Sigma_M / d \Omega = K/q^n$ for the textured and isotropic $\mathrm{Nd}_2\mathrm{Fe}_{14}\mathrm{B}/\alpha$-$\mathrm{Fe}$ nanocomposite. $d \Sigma_M / d \Omega$ has been obtained by subtracting, respectively, the total $d \Sigma / d \Omega$ at $9.5 \, \mathrm{T}$; the fits were restricted to the interval $0.4 \, \mathrm{nm}^{-1} \lesssim q \lesssim 0.6 \, \mathrm{nm}^{-1}$. Dotted horizontal line ($n = 4$) corresponds to scattering due to sharp interfaces (Porod) or to exponentially correlated magnetization fluctuations.}
\end{figure*} 

While the textured nanocomposite reveals a power-law type scattering over most of the $q$-range, the isotropic sample exhibits a more structured $d \Sigma / d \Omega$ with significant curvature at lower and medium $q$. This difference in $d \Sigma / d \Omega$ is most likely related to the difference in the average grain sizes and the ensuing magnetization fluctuations on a nanometer length scale: the isotropic sample has an average grain size of $\sim 20 \, \mathrm{nm}$, while the textured Nd-Fe-B possesses a larger particle size of the order of $100 \,\mathrm{nm}$ (compare the TEM images in Fig.~\ref{fig3}). We also note that the $d \Sigma / d \Omega$ of both samples (data not shown) as well as the spin-misalignment SANS cross section $d \Sigma_M / d \Omega$ [Fig.~\ref{fig6}(c)] are characterized by power-law exponents $n$ that are larger than $4$. This is in agreement with the notion of spin-misalignment scattering, i.e., scattering due to canted spins with a characteristic magnetic-field-dependent wavelength \cite{michels2014review}. It is also quite obvious from this observation that the corresponding magnetization fluctuations in real space are not exponentially correlated (see Fig.~\ref{fig8} below).

As discussed previously, for $\mathbf{k}_0 \parallel \mathbf{H}_0$, any anisotropy of $d \Sigma / d \Omega$ (or of $d \Sigma_M / d \Omega$) is indicative of an anisotropic microstructure. At magnetic \textit{saturation}, the \textit{total} SANS signal arises from nanoscale spatial fluctuations in the nuclear density and in the saturation magnetization $M_s(\mathbf{r})$, presumably at internal $\mathrm{Nd}_2\mathrm{Fe}_{14}\mathrm{B}/\alpha$-$\mathrm{Fe}$ interfaces. The nuclear scattering-length density contrast between the $\mathrm{Nd}_2\mathrm{Fe}_{14}\mathrm{B}$ phase and the $\alpha$-Fe phase amounts to $\Delta \rho_{\mathrm{nuc}} \cong 1.63 \times 10^{14} \, \mathrm{m}^{-2}$, whereas -- at saturation -- the magnetic contrast can be estimated as $\Delta \rho_{\mathrm{mag}} = b_H \Delta M \cong 1.37 \times 10^{14} \, \mathrm{m}^{-2}$, where $\Delta M$ denotes the difference in saturation magnetization between $\alpha$-Fe ($J_s = 2.2 \, \mathrm{T}$) and $\mathrm{Nd}_2\mathrm{Fe}_{14}\mathrm{B}$ ($J_s = 1.61 \, \mathrm{T}$). By assuming that the elements of the microstructure which give rise to nuclear scattering $|\widetilde{N}|^2$ are identical to those which give rise to longitudinal magnetic scattering $b_H^2 |\widetilde{M}_z|^2$, one finds for a saturated sample that the ratio of nuclear to magnetic SANS equals $|\widetilde{N}|^2/(b_H^2 |\widetilde{M}_z|^2) \cong 1.42$. With reference to the electron-microscopy results (Fig.~\ref{fig3}), which reveal a (weakly) anisotropic grain shape (aspect ratio $\sim 1.3$), it is then obvious that a (weakly) horizontally-elongated SANS pattern can already be generated \textit{at saturation} by the combined action of the nuclear and longitudinal magnetic form factors.

\begin{figure}
\includegraphics[width=1.0\columnwidth]{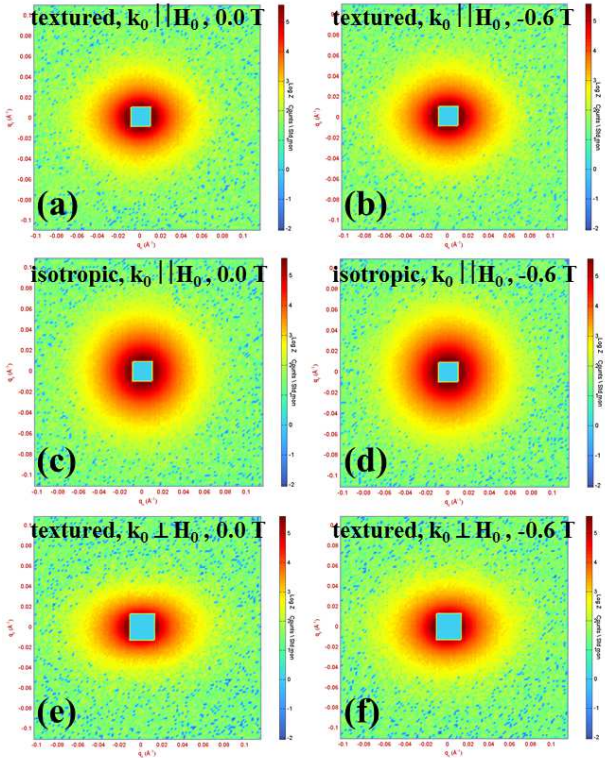}
\caption{\label{fig7} Selected results for the spin-misalignment SANS cross section $d \Sigma_M / d \Omega$ of the textured and isotropic $\mathrm{Nd}_2\mathrm{Fe}_{14}\mathrm{B}/\alpha$-$\mathrm{Fe}$ nanocomposite for $\mathbf{k}_0 \parallel \mathbf{H}_0$ [(a)$-$(d)] and for $\mathbf{k}_0 \perp \mathbf{H}_0$ [(e)$-$(f)] (logarithmic color scale). The respective data set at the maximum applied field of $9.5 \, \mathrm{T}$ has been subtracted. In (a)$-$(d), $\mathbf{H}_0$ is normal to the detector plane, whereas in (e)$-$(f) $\mathbf{H}_0$ is horizontal in the plane.}
\end{figure}

Subtracting the total $d \Sigma / d \Omega$ at $9.5 \, \mathrm{T}$ from the total $d \Sigma / d \Omega$ at lower fields, we obtain the spin-misalignment SANS cross section $d \Sigma_M / d \Omega$ [Eq.~(\ref{sigmasmpara})], which is free of nuclear SANS. The results for $d \Sigma_M / d \Omega$ for the textured nanocomposite [Fig.~\ref{fig7}(a) and \ref{fig7}(b)] still reveal an angular anisotropy with maxima parallel and antiparallel to the horizontal texture axis. Inspection of Eq.~(\ref{sigmasmpara}) then suggests that this observation may be due to (i) spin components which are directed along the $\pm \mathbf{e}_y$-direction [cf.\ the term $\Delta |\widetilde{M}_y|^2 \cos^2\theta$ in Eq.~(\ref{sigmasmpara})] and/or due to (ii) the particle form factor anisotropy (cf.\ terms $\propto \Delta |\widetilde{M}_z|^2$). However, measurements in the $\mathbf{k}_0 \perp \mathbf{H}_0$ geometry [compare Fig.~\ref{fig1}(a)] suggest that longitudinal magnetization fluctuations play only a minor role: if the $d \Sigma_M / d \Omega$ (for $\mathbf{k}_0 \perp \mathbf{H}_0$) were dominated by $\Delta |\widetilde{M}_z|^2$, a $\sin^2\theta$-type anisotropy with intensity maxima along the vertical direction would result [compare Eq.~(\ref{sigmatotperp})]. This is, however, not visible in the experimental data [Fig.~\ref{fig7}(e) and \ref{fig7}(f)], which exhibit a horizontal elongation [cf.\ the term $|\widetilde{M}_y|^2 \cos^2\theta$ in Eq.~(\ref{sigmatotperp})]. In other words, the anisotropy of the scattering pattern for $\mathbf{k}_0 \parallel \mathbf{H}_0$ [Fig.~\ref{fig7}(a) and \ref{fig7}(b)] is due to an anisotropy in the magnetic microstructure, not to the form-factor anisotropy $\Delta |\widetilde{M}_z|^2$ of the particles.

We emphasize that, although the mean magnetization in the remanent state is directed along the $c$-axis [$\mathbf{e}_z$-direction in Fig.~\ref{fig1}(a) and $\mathbf{e}_x$-direction in Fig.~\ref{fig1}(b)], the magnetic neutron scattering cross section is in both geometries dominated by the respective $|\widetilde{M}_y|^2 \cos^2\theta$ term, which (in real space) is related to small misaligned spin components varying along the $\pm \mathbf{e}_y$-direction \cite{michels03epl}. This anisotropy in the spin microstructure may be related to the finding of a crystallographic texture: as shown in Fig.~\ref{fig4}, diffraction peaks of the type $(h k 0)$ exhibit two maxima along the vertical direction ($\theta = 90^{\circ}$ and $\theta = 270^{\circ}$). The investigation of the relation between this crystallographic texture and the spin texture is of interest in its own right and beyond the scope of this paper. However, we would like to emphasize that recent electron-microscopy and three-dimensional atom-probe tomography work by Liu \textit{et al.} \cite{liu2014} also reports anisotropic properties of the grain-boundary phase in hot-deformed nanocrystalline Nd-Fe-B magnets; namely, these authors found that the concentration of rare-earth elements is higher for intergranular phases parallel to the flat surface of the platelet-shaped $\mathrm{Nd}_2\mathrm{Fe}_{14}\mathrm{B}$ grains as compared to intergranular phases along the short side of the platelets.

The characteristic size of the spin inhomogeneities in the remanent state along the vertical and horizontal direction has been estimated by computing [using Eq.~(\ref{corrfunc})] the correlation function $c(r)$ of the spin misalignment (Fig.~\ref{fig8}). The $\exp(-1)$-lengths are $l_C \cong 53 \, \mathrm{nm}$ along the vertical direction, and $l_C \cong 42 \, \mathrm{nm}$ along the horizontal direction; $l_C \cong 28 \, \mathrm{nm}$ for the isotropic nanocomposite. Note that taking the $\exp(-1)$-lengths does not imply that the correlations decay exponentially. For the textured specimen, both $l_C$ values are smaller than the average particle size, which suggests the existence of intraparticle spin disorder, whereas $l_C \cong D $ for the isotropic sample. Compatible with \cite{liu2014}, these results indicate that the microscopic nature of the microstructural defects (e.g., the $\mathrm{Nd}_2\mathrm{Fe}_{14}\mathrm{B}/\alpha$-$\mathrm{Fe}$ interfaces) along these two directions are different (as is manifest by the different correlation lengths). In this respect field-dependent SANS measurements are helpful, since they allow one to determine the field evolution of $l_C$, from which the size of the defect (causing the spin perturbation) and the exchange correlation length may be obtained \cite{bickapl2013,perigo2015}.

\begin{figure}[tb!]
\includegraphics[width=0.80\columnwidth]{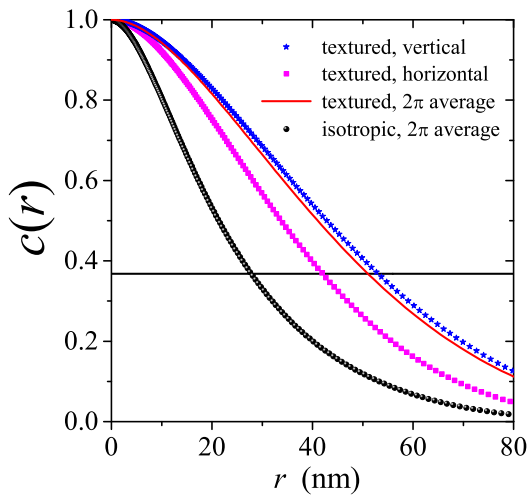}
\caption{\label{fig8} Normalized correlation function $c(r)$ of the spin misalignment [Eq.~(\ref{corrfunc})] for the textured and isotropic $\mathrm{Nd}_2\mathrm{Fe}_{14}\mathrm{B}/\alpha$-$\mathrm{Fe}$ nanocomposite in the remanent state. $c(r)$ of the textured sample has been computed using $d \Sigma_M / d \Omega$ averaged along the vertical and horizontal directions ($\pm 7.5^{\circ}$ sector averages) as well as using the full circular ($2\pi$) average of $d \Sigma_M / d \Omega$; the $c(r)$ of the isotropic sample was computed using the corresponding $2\pi$-averaged $d \Sigma_M / d \Omega$ (see inset). Solid horizontal line: $C(r) = \exp(-1)$. The physically relevant information content of $c(r)$ is restricted to the interval $[r_{\mathrm{min}}, r_{\mathrm{max}}]$ with approximately $r_{\mathrm{min}} \cong 2\pi/q_{\mathrm{max}} = 2 \, \mathrm{nm}$ and $r_{\mathrm{max}} \cong 2\pi/q_{\mathrm{min}} = 130 \, \mathrm{nm}$.}
\end{figure}

\section{Conclusion}

Using magnetic small-angle neutron scattering (SANS) we have provided a comparative study of the magnetic microstructure of textured and isotropic $\mathrm{Nd}_2\mathrm{Fe}_{14}\mathrm{B}/\alpha$-$\mathrm{Fe}$ nanocomposites. Our neutron-data analysis suggests that the spin-misalignment scattering of the textured sample is dominated by spin components along one direction perpendicular to the easy $c$-axis (pressing direction) of the $\mathrm{Nd}_2\mathrm{Fe}_{14}\mathrm{B}$ grains. This anisotropy in the magnetization distribution is accompanied by the presence of a crystallographic texture along these directions. Possible origins for the spin canting (on an intraparticle length scale) have been discussed and are related to the presence of perturbed interface regions, crystalline imperfections, and/or a dispersion in the orientation distribution of the easy $c$-axes. In agreement with the x-ray synchrotron and neutron data, we find anisotropic real-space correlations, with a correlation length that has been estimated at about $40-50 \, \mathrm{nm}$ in the remanent state. The results demonstrate the power of magnetic SANS for analyzing anisotropic magnetic structures on a nanometer length scale; in particular, the complimentary use of the perpendicular and parallel scattering geometries has (for the textured sample) provided results that were otherwise not accessible with only one geometry.

\section*{Acknowledgements}
 
Denis Mettus acknowledges financial support from the National Research Fund of Luxembourg (INTER/DFG/12/07). This paper is based on results obtained from the future pioneering program ``Development of magnetic material technology for high-efficiency motors'' commissioned by the New Energy and Industrial Technology Development Organization (NEDO). The neutron experiments were performed at the Swiss spallation neutron source SINQ, Paul Scherrer Institute, Villigen, Switzerland. ALBA synchrotron is acknowledged for the provision of beamtime. We thank Birgit Heiland (INM, Saarbr\"ucken) and J\"org Schmauch (Universit\"at des Saarlandes) for the electron-microscopy work.

\bibliographystyle{apsrev4-1}

%

\end{document}